\documentclass{elsart}

\usepackage{graphicx}

\usepackage{amssymb}

\journal{Solid State Communications}

\begin{document}

\begin{frontmatter}



\title{Superconductivity in Compressed Potassium and Rubidium}


\author{Lei Shi}
\address{School of Computational Sciences, George Mason University,
Fairfax VA, USA 22032}
\ead{lshi@scs.gmu.edu}

\author{Dimitrios A. Papaconstantopoulos\corauthref{cor}},
\corauth[cor]{Corresponding author.}
\ead{papacon@dave.nrl.navy.mil}
\author{Michael J. Mehl}
\ead{mehl@dave.nrl.navy.mil}
\address{Center for Computational Materials Science, Naval Research
Laboratory, Washington DC, USA 20375-5000}

\begin{abstract}
Calculations of the electron-phonon interaction in the alkali
metals, Potassium and Rubidium, using the results of band theory and
BCS theory-based techniques suggest that at high pressures K and Rb
would be superconductors with transition temperatures approaching
$10 K$.
\end{abstract}

\end{frontmatter}


In a recent paper Shimizu {\em et al.}\cite{shimizu02:super_li}
reported the discovery of superconductivity in compressed Lithium
with a transition temperature $T_{c} = 20K$.  This report is a
confirmation of previous theoretical work of Neaton and
Ashcroft,\cite{neaton99:dense_li} who predicted that at high
pressures Lithium forms a paired ground state, and of Christensen
and Novikov,\cite{christensen01:super_li} who suggested that fcc
Lithium under increased pressure may reach $T_{c} = 50$-$70K$.

In this work we applied a methodology similar to that of
Ref.\cite{christensen01:super_li} to the alkali metals K and Rb.
The procedure goes as follows: we first performed Augmented Plane
Wave (APW) calculations of the band structure and total energy of
the above alkali metals in both the bcc and fcc structures over a
wide range of volumes reaching high pressures.  From these
calculations we obtained the Fermi level, $\varepsilon_{F}$, values
of the density of states, $N_{t}$, and its angular momentum
decomposition, $N_{l}$, as a function of volume.  We also used the
APW results to determine the volume variation of the bulk modulus B.
We then used the self-consistent APW potentials to determine the
scattering phase shifts $\delta_{l}$ again as a function of volume.
The quantities $N_{l}$ and $\delta_{l}$ were then used in the
``rigid muffin-tin''
approximation\cite{gaspari72:lambda,papacon77:super} to determine
the Hopfield parameter $\eta$.  The next step was to calculate the
electron-phonon interaction parameter,
\begin{equation}
\lambda = \frac{\eta}{M<\!\!\omega^2\!\!>} ~ .
\label{equ:epi}
\end{equation}
We accomplished this by assuming\cite{jarlborg88} that
\begin{equation}
<\omega^2> = C B V^{1/3} ~ ,
\label{equ:omeg2}
\end{equation}
where the constant of proportionality C was determined from from the
experimental values of B, V and the Debye temperature $\theta_D$.

\begin{figure}
\includegraphics*[width=4in]{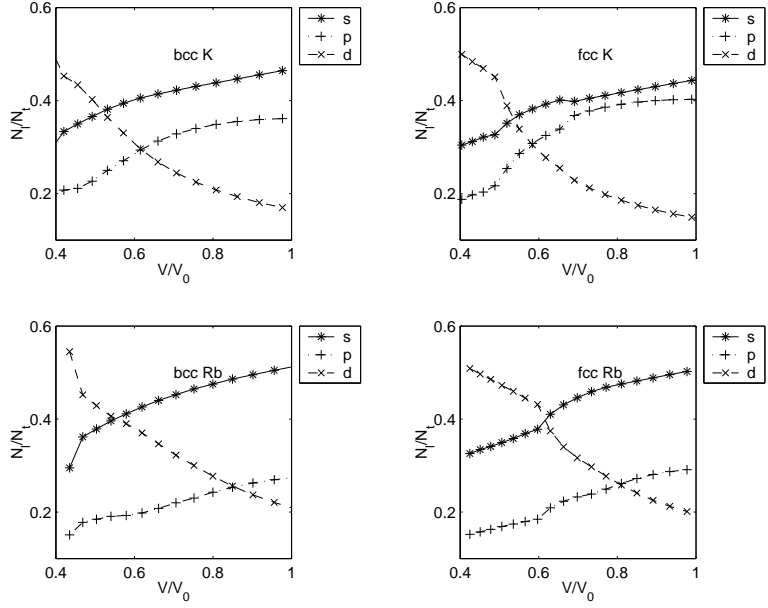}
\caption{Angular Momentum decomposed DOS divided by total DOS at
$\varepsilon_{F}$}
\label{fig:spd}
\end{figure}

\begin{figure}
\includegraphics*[width=4in]{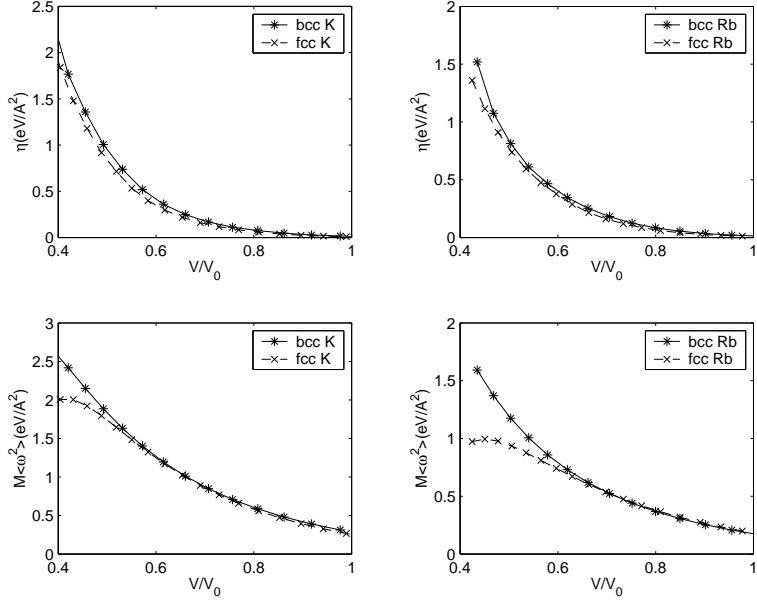}
\caption{ $\eta$ and $M<\!\!\omega^2\!\!>$ as a function of volume.}
\label{fig:etaomega}
\end{figure}

In Fig.~\ref{fig:spd} we show the ratios
$N_{l}(\varepsilon_{F})/N_{t}(\varepsilon_{F})$ at the Fermi level
$\varepsilon_{F}$ as a function of volume.  These ratios are crucial
in the determination of $\eta$.  It is important to note that the
ratio $N_{d}(\varepsilon_{F})/N_{t}(\varepsilon_{F})$ increases
rapidly as we go to smaller volumes.  This build up of the d-like
DOS under pressure causes the large values of $\eta$ at small
volumes shown in Fig.~\ref{fig:etaomega}.

In this theory there are three contributions to $\eta$ coming from
the channels s-p, p-d and d-f.  The d-f contribution is negligible
but the p-d contribution is comparable to the s-p at large volumes
and becomes larger at small volumes where superconductivity occurs.
Also in Fig.~\ref{fig:etaomega}, we show the variation with volume
of the denominator of (\ref{equ:epi}), $M<\!\!\omega^2\!\!>$,which
we have extracted from the variation of the bulk modulus.

\begin{figure}
\includegraphics*[width=4in]{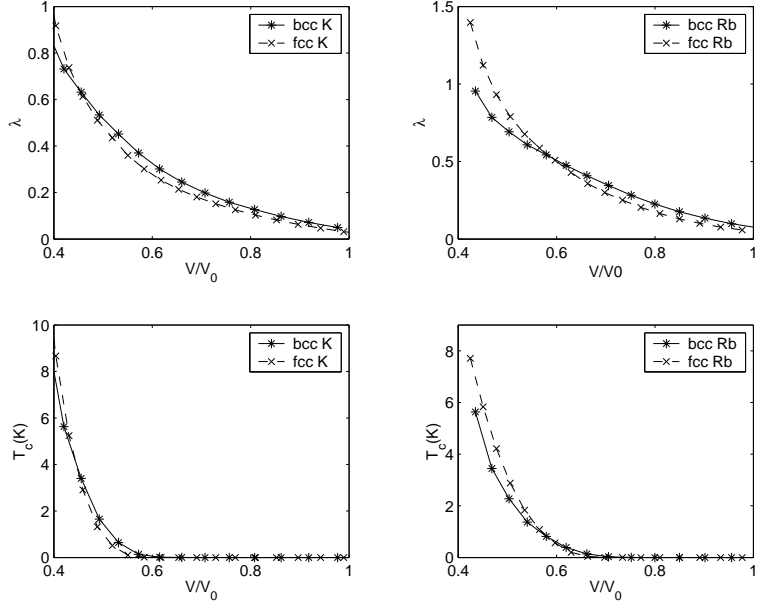}
\caption{ $\lambda$ and $T_{c}$ as function of volume.}
\label{fig:lamdatc}
\end{figure}

In Fig.~\ref{fig:lamdatc}, we show the electron-phonon coupling
$\lambda$ as a function of volume determined from (\ref{equ:epi}).
It is evident that at small volumes $\lambda$ reaches large values
suggesting that these metals can display superconductivity under
pressure.  To quantify our prediction for superconductivity on the
basis of strong electron-phonon coupling, we have calculated $T_{c}$
using the McMillan equation\cite{mcmillan68:tc,allen75:tc} with a
Coulomb pseudopotential value $\mu^*=0.13$.  These results are also
shown in Fig.~\ref{fig:lamdatc}.

It is clear that transition temperatures in a range of 5-10~K are
reachable for both the fcc and bcc lattices at volumes in the
neighborhood of $V/V_{0} \approx 0.4$.  We have calculated that such
volumes correspond to pressures of 13.5~GPa for K and 8~GPa for Rb,
respectively.

The similarity of our fcc and bcc results suggests that our
prediction of superconductivity in K and Rb is independent of
crystal structure.  We believe that our prediction is still valid
even if, experimentally, these materials under high pressure
transform to other structures such as \emph{hR}1 or
\emph{cI}16.\cite{hanfland00:hiprs_li} We propose that the mechanism
of superconductivity in these metals is due to the increased d-like
character of the wave-functions at $\varepsilon_{F}$ at high
pressures, which validates our use of the ``Rigid Muffin-tin''
approximation that is successful in transition metals.



\end{document}